# Final state interactions and NN correlations: are the latter measurable?

A.Bianconi[1], S.Jeschonnek[2], N.N.Nikolaev[2,3], J.Speth[2], B.G.Zakharov[3]

[1] Dipartimento di Fisica Nucleare e Teorica, Università di Pavia, and
Istituto Nazionale di Fisica Nucleare, Sezione di Pavia, Pavia, Italy

[2] IKP(Theorie), Forschungszentrum Jülich GmbH., D-52425 Jülich, Germany

[3] L.D.Landau Institute for Theoretical Physics,
GSP-1, 117940, ul.Kosygina 2, V-334 Moscow, Russia

## Abstract

Are effects of short range correlations in a ground state of the target nucleus (initial state correlations ISC) observable in experiments on quasielastic $A(e,e'p)$ scattering at large missing momentum $p_m$? Will the missing momentum spectrum observed at CEBAF be overwhelmed by final state interactions (FSI) of the struck proton? The recent advances in the theory of FSI and findings of complex interplay and strong quantum-mechanical interference of FSI and ISC contributions to scattering at $p_m \gtrsim 1\,\mathrm{fm}^{-1}$ are reviewed. We conclude that for $p_m \gtrsim 1\ \mathrm{fm}^{-1}$ quasielastic scattering is dominated by FSI effects and the sensitivity to details of the nuclear ground state is lost.

## 1 Introduction

Ever since Gottfried's classic theoretical works ([1], see also [2]), an investigation of short range ISC in nuclei is considered one of the principal goals of experiments on quasielastic $A(e,e'p)$ scattering at large missing momentum $p_m$ (for reviews see [3]). Such experiments are becoming a reality at a new generation of high luminosity, continuous beam, electron facilities (CEBAF, AmPS, MAMI, Bates). Especially important is a new domain of GeV energies attainable at CEBAF. The experimentally measured $p_m$ distribution is distorted by FSI of the struck proton in the target nucleus debris. Interpretation of the experimental data taken at non-relativistic energies of the struck proton is plagued by rapid energy dependence of FSI. At CEBAF for the first time the missing energy–momentum spectrum will be exhausted under conditions of negligible variation of FSI parameters.

Is FSI very strong or just leads to small corrections to the ISC contribution to large-$p_m$ phenomena? The recent work [4, 5, 6, 7, 8, 9, 10, 11] on the theory of FSI at high $Q^2$ called in question the mere possibility of the theoretical interpretation of the experimental data on large $p_m$ in terms of the ground state correlations. In this talk we review this new important development in the theory of FSI at GeV energies.

From the technical point of view, at large $Q^2 \gtrsim (1-2)\,GeV^2$ and large $T_{kin} \approx Q^2/2m_p$, the very description of FSI becomes substantially different from the conventionally used non-relativistic (optical) potential model DWIA. Namely, the nature of nucleon-nucleon interaction changes from the purely elastic, potential scattering at low energies to a diffractive, strongly absorptive, small angle scattering at $T_{kin} \gtrsim (0.5-1)\,GeV$. In this diffractive regime the potential approach breaks down and Glauber's multiple scattering theory [12] becomes a natural framework for quantitative description of FSI and leads to several important new effects in the calculation of FSI-modified one-body density matrix and missing momentum distribution in $A(e,e'p)$ scattering, which are missed in the conventional DWIA.

As a matter of fact, a very strong absorption of struck protons in the target nucleus debris is a well established fact: it reduces the observed cross section almost twofold for carbon and fourfold for the gold nucleus [13] in perfect agreement with the Glauber theory predictions [5]. Little wonder that such a strong attenuation of the struck proton wave is accompanied by a strong distortion of the missing momentum distribution and these distortions are especially strong at large $p_m$. The quantitative description of these distortion effects requires a careful quantum mechanical treatment of the interaction dynamics, which is reviewed in this talk.

## 2   Missing momentum distribution: kinematics and definitions

The $A(e,e'p)$ coincidence cross section can be represented in the form [3, 14]

$$\frac{d\sigma}{dQ^2 d\nu dp d\Omega_p} = K|M_{ep}|^2 S(E_m, \vec{p}_m, \vec{p}) . \tag{1}$$

Here $K$ is a kinematical factor, $M_{ep}$ is the $ep$ elastic scattering amplitude, $\nu$ and $\vec{q}$ are the $(e,e')$ energy and momentum transfer, $Q^2 = \vec{q}^{\,2} - \nu^2$, the struck proton has a momentum $\vec{p}$ and energy $E(p) = T_{kin} + m_p$, the missing momentum and energy are defined as $\vec{p}_m = \vec{q} - \vec{p}$ and $E_m = \nu - T_{kin} - T_{kin}(A-1)$ (where $T_{kin}(A-1)$ is the kinetic energy of the undetected (A-1) residual system) and the $z$-axis is chosen along $\vec{q}$.

Clean signal of FSI effects is seen in the inclusive missing momentum spectrum

$$W(\vec{p}_m) = \frac{1}{(2\pi)^3} \int dE_m S(E_m, \vec{p}_m) = \frac{1}{(2\pi)^3} \int d\vec{r}\,' d\vec{r}\, \rho(\vec{r},\vec{r}\,') \exp\left[i\vec{p}_m(\vec{r}-\vec{r}\,')\right] , \tag{2}$$

where

$$\rho(\vec{r},\vec{r}\,') = \int d\{\vec{R}\}\, \Psi^*(\{\vec{R}\},\vec{r}\,')\hat{S}^\dagger(\{\vec{R}\},\vec{r}\,')\hat{S}(\{\vec{R}\},\vec{r})\Psi(\{\vec{R}\},\vec{r}) \tag{3}$$

is the FSI-modified one-body density matrix (OBDM). Here $\vec{r} = \vec{r}_A$ is the coordinate of the struck proton, $\{\vec{R}\} = \{\vec{r}_1, ...., \vec{r}_{A-1}\}$ stands for coordinates of the spectator protons (for the sake of brevity, we suppress the c.m.s motion constraints, the full machinery of the Jacobi coordinates has been used in all the practical calculations), $\hat{S}(\{\vec{R}\},\vec{r})$ describes FSI of the struck proton with the spectator nucleons. In the PWIA, when $\hat{S} = 1$, eq. (3) reduces to the standard OBDM matrix for a ground state of a nucleus and eq. (2) gives the familiar SPMD $N(p_m)$, often referred to as the PWIA momentum distribution.

We focus on an evaluation of the OBDM (3) and missing momentum distribution $W(\vec{p}_m)$. In $D(e,e'p)$ scattering, one can directly use the deuteron wave function from realistic potential models. The real testing ground for the comparison of FSI and ISC effects is the $^4He(e,e'p)$ scattering. Here one can perform an exhaustive analysis of the interplay of FSI and ISC effects, making use of the standard Jastrow correlated wave function $\Psi(\vec{r}_1,...,\vec{r}_A) \equiv \Psi_o(\vec{r}_1,....,\vec{r}_A)\hat{F}$, where the mean field wave function $\Psi_o(\vec{r}_1,....,\vec{r}_A)$ can be approximated by the harmonic oscillator wave function, $\hat{F} \equiv \prod_{i<j}^{4}\left[1 - C(\vec{r}_i - \vec{r}_j)\right]$, and $C(r) = C_o \exp\left(-r^2/2r_c^2\right)$. For a hard core repulsion $C_o = 1$, for a soft core $C_o < 1$. One usually considers $r_c \approx 0.5\text{-}0.6$ fm [15].

## 3   Final state interaction and the Glauber theory

For the reason of strongly absorptive $NN$ interaction, at large $Q^2$ and high $T_{kin} \gtrsim (0.5\text{-}1)\,\text{GeV}$ the familiar nonrelativistic potential approach to FSI breaks down. However, the wavelength of the struck proton is short compared to the size of the nucleon and internucleon separation and the Glauber theory becomes applicable. Defining $\vec{r}_i \equiv (\vec{b}_i, z_i)$, we can write

$$\hat{S}(\{\vec{R}\}, \vec{r}) = \prod_{i=1}^{A-1} \left[ 1 - \theta(z_i - z)\Gamma(\vec{b} - \vec{b}_i) \right]. \tag{4}$$

The $\theta$-function in (4) tells that the struck proton can have FSI only with the forward hemisphere spectators. The standard parameterization of the $pN$ profile function is

$$\Gamma(\vec{b}) = \frac{\sigma_{tot}(1 - i\rho)}{4\pi b_o^2} \exp\left[ -\frac{\vec{b}^2}{2b_o^2} \right] \tag{5}$$

Here $\rho$ is the $Re/Im$ ratio for the forward elastic $pN$ scattering amplitude and $b_0$ is the $pN$ interaction radius ($b_0^2$ is the diffraction slope). The Glauber theory has no free parameters. The long history of its successes is well documented [16]. At $T_{kin} \sim 1$GeV, the $pN$ scattering data give $b_o \approx 0.5 fm$, $\sigma_{tot} \approx 40 mb$ and $\rho = .33$ [16, 17]. These FSI parameters only weakly vary over the GeV energy range of CEBAF. This makes FSI insensitive to the missing energy $E_m$. We have implicitly used this property in (2,3). Notice a proximity of the FSI and ISC radii: $b_0 \approx r_c$.

## 4 Short-ranged and long-ranged components of the FSI-modified one-body density matrix

The FSI-modified OBDM can be written as

$$\rho(\vec{r}, \vec{r}') = \int d\{\vec{R}\} \Psi_o^*(\{\vec{R}\}, \vec{r}') \hat{F}^\dagger(\{\vec{R}\}, \vec{r}') \hat{S}^\dagger(\{\vec{R}\}, \vec{r}') \hat{S}(\{\vec{R}\}, \vec{r}) \hat{F}(\{\vec{R}\}, \vec{r}) \Psi_o(\{\vec{R}\}, \vec{r}) \tag{6}$$

The operator $\hat{F}^\dagger \hat{S}^\dagger \hat{S} \hat{F}$ which emerges in (6) can be expanded as [7]

$$\hat{F}^\dagger \hat{S}^\dagger \hat{S} \hat{F} = \prod_{i<j}^{A} \left[ 1 - C^\dagger(\vec{r}_i{'} - \vec{r}_j{'}) \right] \left[ 1 - C(\vec{r}_i - \vec{r}_j) \right]$$

$$\times \prod_{i}^{A-1} \left[ 1 - \theta(z_i - z')\Gamma^\dagger(\vec{b}' - \vec{b}_i) \right] \left[ 1 - \theta(z_i - z)\Gamma(\vec{b} - \vec{b}_i) \right] =$$

$$1 - \sum \left[ C^\dagger + C \right] - \sum \left[ \Gamma^\dagger + \Gamma \right] + \sum [C^\dagger \Gamma + C\Gamma^\dagger] + \sum C^\dagger C + \sum \Gamma^\dagger \Gamma + \dots \tag{7}$$

In this cluster expansion, the higher order terms in $C, C^\dagger, \Gamma, \Gamma^\dagger$ can be considered as "interactions" which modify the OBDM $\rho_0(\vec{r}, \vec{r}')$ of the mean field approximation. There are interactions which only involve one of the trajectories in the FSI-modified OBDM, there are terms of the form $C^\dagger(\vec{r}' - \vec{r}_i{'})C(\vec{r} - \vec{r}_i)$, $\theta(z_i{'} - z')\theta(z_i - z)\Gamma^\dagger(\vec{b}' - \vec{b}_i{'})\Gamma(\vec{b} - \vec{b}_i)$ and $C^\dagger(\vec{b}' - \vec{b}_i)\Gamma(\vec{b}_4 - \vec{b}_i)$+h.c., which lead to a short-ranged (on the scale $r_c$ and/or $b_0$, as opposed to a large nuclear radius) interaction between the two trajectories in the calculation of OBDM and govern the large $p_m$ momentum spectrum [4, 5, 6, 7], there is a novel effect [7] of quantum-mechanical ISC-FSI interference due to the $C^\dagger \Gamma + \Gamma^\dagger C$ interactions etc.

In order to set up a background, we recall the salient features of SPMD $N(\vec{p}_m)$ for the example of the $^4He$ nucleus [7, 11]. In the mean field approximation, one finds a long ranged OBDM $\rho_0(\vec{r}, \vec{r}') \propto \exp\left(-\frac{3}{8R_o^2}\left(\vec{r}^2 + \vec{r}'^2\right)\right)$ and steeply decreasing SPMD

$$N(1; \vec{p}_m) = w_1 \exp\left(-\frac{4}{3}R_o^2 p_m^2\right) \tag{8}$$

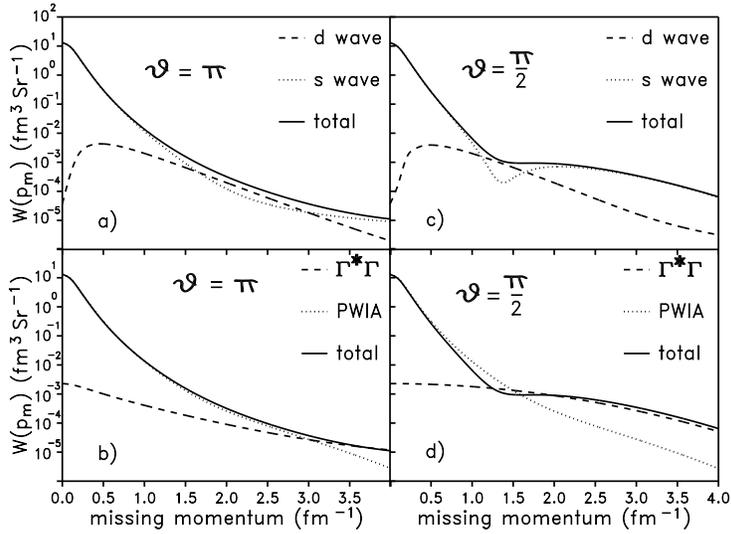

Figure 1: The predicted FSI-modified missing momentum distribution (solid curve) $W(\vec{p}_m)$ for $D(e,e'p)$ scattering [8]. Shown separately are (a,c) the $S$ and $D$-wave contributions and (b,d) the PWIA and the $\propto \Gamma^*\Gamma$ rescattering contributions at (a,b) $\theta = 180°$ and (c,d) $\theta = 90°$.

with the Fermi momentum $k_F \sim 1/R_0$. (Hereafter $N(1;\vec{p}_m)$, $N(C;\vec{p}_m)$,... indicate the contributions to $N(\vec{p}_m)$ coming from the "1","C",... terms in the expansion (7)). The leading corrections to $N(\vec{p}_m)$ come from interactions linear in $C(\vec{r}-\vec{r}_j), C^\dagger(\vec{r}'-\vec{r}_j)$, which affect only one of the trajectories in the calculation of the OBDM:

$$N(C^\dagger + C;\vec{p}_m) \approx -6w_1 C_0 \sqrt{\frac{27}{125}} \left(\frac{r_c}{R_o}\right)^3 \exp\left(-\frac{4}{5}R_o^2 p_m^2\right). \qquad (9)$$

(For brevity, hereafter we suppress the correction factors $[1+\mathcal{O}(r_c^2/R_o^2)]$ to the slope and the normalization). Notice the small, $\propto (r_c/R_0)^3$, normalization in (9) and the destructive interference between the $N(1;\vec{p}_m)$ and $N(C^\dagger + C;\vec{p}_m)$, which becomes stronger with increasing $p_m$, because the latter has a smaller slope of the $\vec{p}_m^2$ dependence than the former. The driving contribution to the short ranged component of the OBDM and the related large-$p_m$ component of the SPMD, comes from the $C^\dagger(\vec{r}'-\vec{r}_i')C(\vec{r}-\vec{r}_i)$ terms in (7),

$$N(C^\dagger C;\vec{p}_m) \approx w_1 \frac{1}{\pi^3} \frac{1}{R_o^3} \sqrt{\frac{3^5}{2^{15}}} \left|\int d^3\vec{r}\, C(\vec{r})\exp(i\vec{p}_m\vec{r})\right|^2 = w_1 C_o^2 \sqrt{\frac{243}{512}} \left(\frac{r_c}{R_o}\right)^6 \exp\left(-r_c^2 p_m^2\right). \qquad (10)$$

Notice a close semblance of Eq. (10) to the momentum distribution in the deuteron with the short-range correlation function $C(\vec{r})$ playing the rôle of the wave function (see below Eq. (11)), which is the basis of the familiar quasi-deuteron interpretation of the ISC contribution to SPMD.

The fundamental issue is whether the ISC component (10) of SPMD survives the FSI distortions or not and whether the correlation function $C(\vec{r})$ can be measured in $A(e,e'p)$ scattering or not.

## 5 Final state interaction effects in $D(e,e'p)$ scattering [8,10]

The realistic models [18] of the deuteron wave function allow an accurate evaluation of FSI effects [8]. The quadrupole deformation of the deuteron gives an interesting handle on the strength of FSI in the tensor polarized deuteron [10]. For unpolarized deuterons and spin-independent $pn$ scattering amplitude

$$W(\vec{p}_m) = \frac{1}{4\pi(2\pi)^3} \int d^3\vec{r}\,d^3\vec{r}\,' \exp[i\vec{p}_m \cdot (\vec{r} - \vec{r}\,')] S(\vec{r}) S^\dagger(\vec{r}\,') \left[ \frac{u(r)}{r}\frac{u(r')}{r'} + \frac{1}{2}\frac{w(r)}{r}\frac{w(r')}{r'}\left(3\frac{(\vec{r}\cdot\vec{r}\,')^2}{(rr')^2} - 1\right) \right], \tag{11}$$

where $u$ and $w$ are the familiar $S$ and $D$ wave functions and $S(\vec{r}) = 1 - \theta(-z)\Gamma(\vec{b})$. The anisotropy of $S(\vec{r})S^\dagger(\vec{r}\,')$ leads to the angular anisotropy of $W(\vec{p}_m)$ shown in Figs. 1,2.

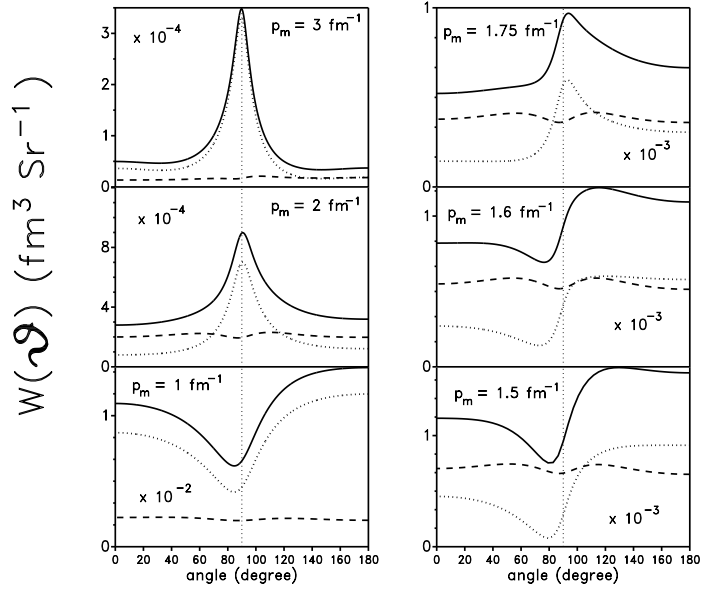

Figure 2: The predicted FSI-modified angular dependence of (solid curve) $W(\vec{p}_m)$ for the deuteron target and of its (dotted curve) $S$-wave and (dashed curve) $D$-wave components at different values of $p_m$ [8].

The forward-backward asymmetry $W(p_\perp, -p_z) \neq W(p_\perp, p_z)$ has its origin in the nonvanishing real part of the $p$-$n$ scattering amplitude $\rho \neq 0$, for which $S(b, z)S^\dagger(b', z') \neq S(b', z')S^\dagger(b, z)$.

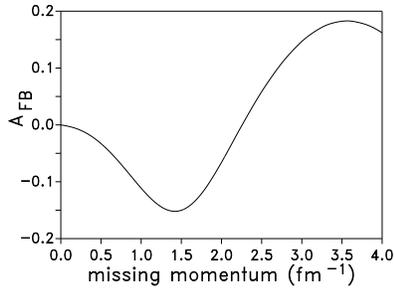

Figure 3: The predicted forward-backward asymmetry $A_{FB}(p_m) = \frac{W(p_m, \theta=0°) - W(p_m, \theta=180°)}{W(p_m, \theta=0°) + W(p_m, \theta=180°)}$ for the unpolarized deuteron [8].

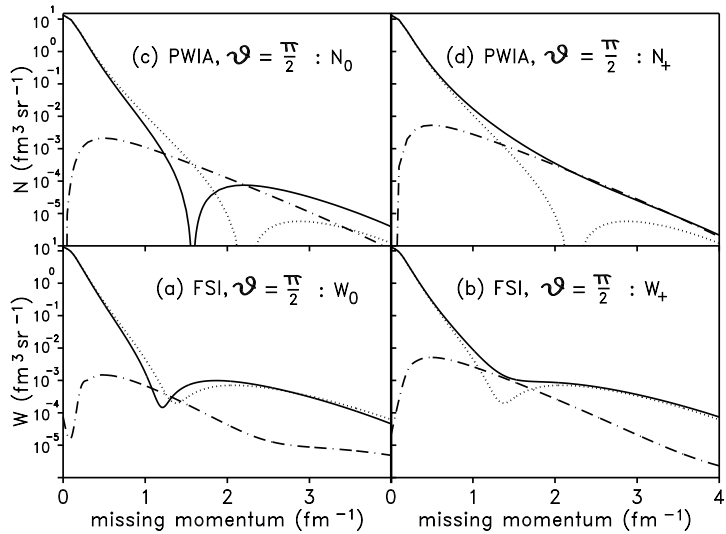

Figure 4: Predictions for the full FSI-modified momentum distributions (solid line) $W_o$ (a), $W_+$ (b) and PWIA distributions $N_o$ (c), $N_+$ (d) vs. $p_m$ at $\theta = 90°$ for the tensor polarized deuteron target [10] . Shown separately are the S-wave (dotted line) and D-wave (dash-dotted line) contributions.

The angular asymmetry mostly comes from FSI distortions of the $S$-wave amplitude which can be decomposed into the PWIA and FSI terms as

$$U(\vec{p}_m) = \int d^3\vec{r}\exp(-\vec{p}_m \cdot \vec{r})S(\vec{r})\frac{u(r)}{r} = u(1;\vec{p}_m) - u(\Gamma;\vec{p}_m) \qquad (12)$$

The PWIA term $u(1;\vec{p}_m)$ decreases on the scale $p_\perp, p_{m,z} \sim 1/R_D$. The FSI term is small, $\propto \sigma_{tot}(pn)/(2\pi R_D^2) \sim 0.035$ [5, 8], but takes over at $p_\perp \gtrsim 1\,\text{fm}^{-1}$, where

$$W(\vec{p}_m) \propto |u(\Gamma;\vec{p}_m)|^2 \propto \frac{d\sigma_{el}(pn \to pn)}{dp_\perp^2} \propto \exp(-b_0^2 p_\perp^2). \qquad (13)$$

In Fig. 1, we show how the destructive interference of the PWIA and FSI amplitudes produces a dip at $p_\perp \sim 1.3\,\text{fm}^{-1}$ in transverse kinematics. Fig. 2 illustrates the angular anisotropy of $W(\vec{p}_m)$. At small $p_m \ll 1\,\text{fm}^{-1}$ the angular distribution is isotropic, with the increasing $p_m$ it first develops an approximately symmetric dip at $\theta = 90°$, which through a very asymmetric stage evolves into the sharp elastic rescattering peak at $90°$ at larger values of $p_m$. Fig. 1d shows the decomposition of the same $90°$ distribution into the PWIA and the FSI components.

Fig. 1b shows that for $p_m \gtrsim 1.5\,\text{fm}^{-1}$, FSI effects are substantial also in parallel kinematics. At large $p_m$, they are entirely due to the "elastic rescattering" operator $\theta(-z)\theta(-z')\Gamma(b)\Gamma^*(b')$ in the integrand of (11). Indeed, $\theta(-z)\theta(-z') = \theta(-z_{max})$, where

$$z_{max} = \frac{1}{2}(z + z') + \frac{1}{2}|z - z'| \qquad (14)$$

and the non-analytic function $|z - z'|$ in $z_{max}$ gives rise to a $\propto p_{m,z}^{-2}$ tail of $W(\vec{p}_m)$ after the Fourier transform in $z - z'$ (see also Section 9). It is worthwhile to mention that the $\theta(z)$ in the Glauber operator assumes idealized pointlike nucleons. The effect of the non-pointlike nucleons can be modelled by a smearing, for instance, $\theta(z) \to \frac{1}{2}[1 + \tanh(\frac{z}{z_o})]$. The educated guess for the smearing width is $2z_0 \sim r_c$. Because the realistic wave functions of the deuteron already include a suppression of small size configurations by a short range repulsion, the uncertainties with the smearing are small at least up to $p_m \lesssim 3\,\mathrm{fm}^{-1}$.

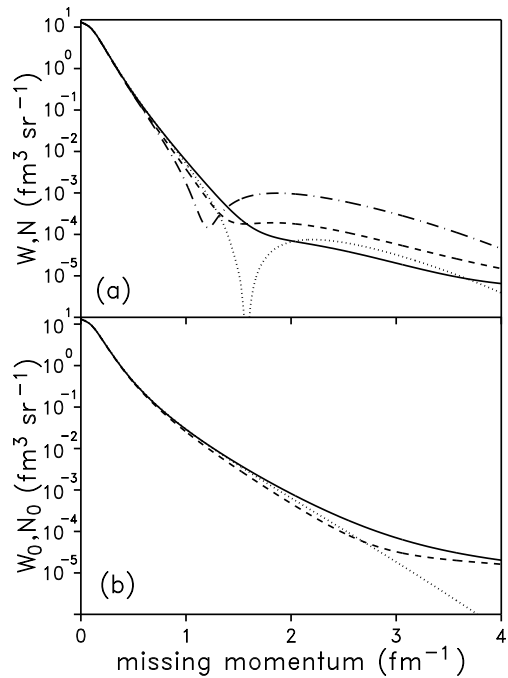

Figure 5: Predictions for the tensor polarized deuteron target [10]: (a) The FSI-modified distributions $W_+(p_m, \theta = 0°)$ (solid line), $W_+(p_m, \theta = 180°)$ (dashed line) and $W_o(p_m, \theta = 90°)$ (dash-dotted line) and the PWIA distributions $N_+(p_m, \theta = 0°) = N_+(p_m, \theta = 180°)$ (dotted line). (b) The FSI-modified distributions $W_o(p_m, \theta = 0°)$ (solid line), $W_o(p_m, \theta = 180°)$ (dashed line) and the PWIA distributions $N_o(p_m, \theta = 0°) = N_o(p_m, \theta = 180°)$ (dotted line).

The asymmetry in parallel kinematics, $A_{FB} = [W(\theta = 0°, p_m) - W(\theta = 180°, p_m]/[W(\theta = 0°, p_m) + W(\theta = 180°, p_m)]$ (Fig. 3), is quite strong. Implications of strong FSI effects in longitudinal kinematics for the $y$-scaling analysis are discussed in [8].

Different models for the deuteron wave function differ mostly in the D-wave contribution [18]. The D-wave controls the quadrupole deformation of the deuteron and hopefully can best be probed in $\vec{D}(e, e'p)$ scattering on tensor polarized deuterons (the spin quantization axis is taken along $\vec{q}$). Here a key observable is the tensor analyzing power $A \equiv (\sigma_+ + \sigma_- - 2\sigma_0)/(\sigma_+ + \sigma_- + \sigma_0)$ where $\sigma_\mu$ is the cross section for the polarization state $\mu$ (notice that $\sigma_+ = \sigma_-$). The interplay of the quadrupole deformation and FSI distortions, which are very different for different D-wave amplitudes, leads to a very rich pattern of FSI effects [10]. The dependence of $W_\mu(\vec{p}_m)$ (as well as of PWIA distributions $N(\vec{p}_m)$) on the polarization state is strong, see Figs. 4,5. Fig. 5a shows very clearly how FSI destroys the PWIA symmetry relation $N_0(p_m, \theta = 90°) = N_\pm(p_m, \theta = 0°, 180°)$. A substantial departure from the PWIA distributions starts at $p_m \gtrsim 1\,\mathrm{fm}^{-1}$. In the PWIA, both $N_\mu(\vec{p}_m)$ and tensor analyzing power $A$ exhibit a sensitivity to the model of the wave function, which is washed out by FSI effects, see Figs. 6,7. In general, FSI caused departures from PWIA predictions are dramatically stronger than the difference between the Bonn and Paris model predictions.

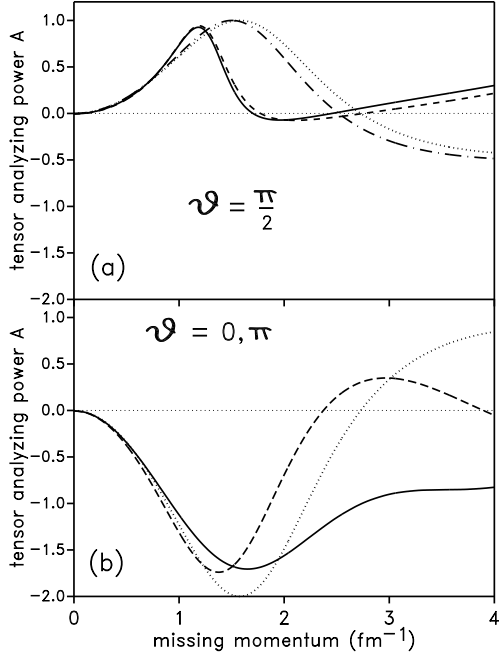

Figure 6: Predictions [10] for the PWIA and FSI-modified tensor analyzing power in $\vec{D}(e,e'p)$ scattering vs. the missing momentum $p_m$. Panel (a) shows $A^{PWIA}(p_m,\theta=90°)$ calculated using the Bonn wave function (dotted line) and calculated using the Paris wave function (dash-dotted line) and also the FSI-modified $A(p_m,\theta=90°)$ calculated with the Bonn wave function (solid line) and with the Paris wave function (dashed line). Panel (b) shows $A^{PWIA}(p_m,\theta=0°) = A^{PWIA}(p_m,\theta=180°)$ (dotted line) and FSI-modified $A$ for $\theta=0°$ (solid line) and $\theta=180°$ (long-dashed line).

# 6  FSI in heavy nuclei: the formalism

Some properties of FSI are best seen in heavy nuclei, neglecting the ISC altogether. In this limit, $\rho(\vec{r},\vec{r}') = \rho_0(\vec{r},\vec{r}')\Phi(\vec{r},\vec{r}')$, where

$$\rho_0(\vec{r},\vec{r}') = \frac{1}{Z}\sum_n \phi_n(\vec{r})\phi_n^*(\vec{r}') \tag{15}$$

is the familiar shell model OBDM, the FSI distortion factor equals

$$\Phi(\vec{r},\vec{r}') = \int \prod_{j=1}^{A-1} \rho_A(\vec{r}_j) d^3\vec{r}_j S^\dagger(\vec{r}_1,...,\vec{r}_{A-1},\vec{r}')S(\vec{r}_1,...,\vec{r}_{A-1},\vec{r}) \tag{16}$$

and $n_A(\vec{r}) = A\rho_A(\vec{r})$ is the nuclear density. The Glauber formula (4) gives

$$\Phi(\vec{r},\vec{r}') = \left[1 - \frac{1}{A}\int d^2\vec{b}_1 \Gamma(\vec{b}-\vec{b}_1)t(\vec{b}_1,z) - \frac{1}{A}\int d^2\vec{b}\Gamma^\dagger(\vec{b}'-\vec{b}_1)t(\vec{b}_1,z') \right.$$
$$\left. + \frac{1}{A}\int d^2\vec{b}_1 \Gamma^\dagger(\vec{b}'-\vec{b}_1)\Gamma(\vec{b}-\vec{b}_1)t(\vec{b}_1,\max(z,z'))\right]^{A-1}, \tag{17}$$

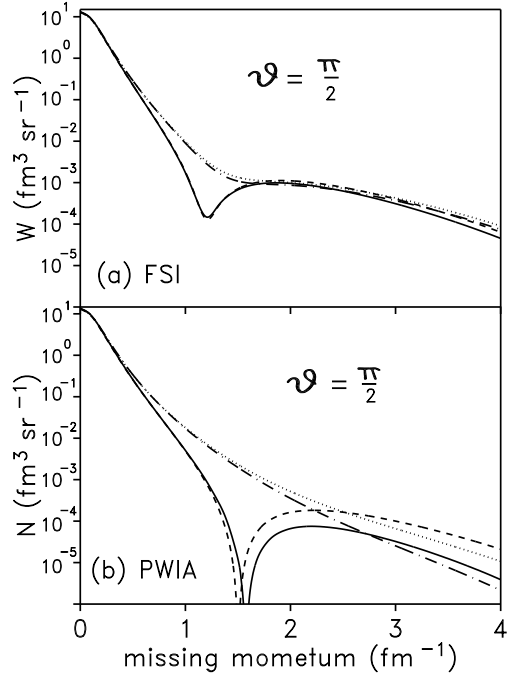

Figure 7: Predicted [10] momentum distributions for the tensor polarized deuteron target calculated with the Bonn and Paris wave functions in PWIA (b) and including FSI (a) at $\theta = 90°$. The results calculated with the Bonn wave function for $W_o$ and $N_o$ are shown by solid lines, $W_+$ and $N_+$ are shown by dash-dotted lines, the results calculated with the Paris wave function for $W_o$ and $N_o$ are shown by the dashed lines and for $W_+$ and $N_+$ are shown by dotted lines.

where $t(\vec{b}, z) = A \int_z^\infty dz_1 \rho_A(\vec{b}, z_1)$ is a partial optical thickness. Because of the $\Gamma^\dagger \Gamma$ interaction between the two trajectories, which is a steep function of $\vec{r} - \vec{r}'$, the FSI factor (17) can not be represented in a factored form in the variables $\vec{r}$ and $\vec{r}'$. Because $t(\vec{b}, z)$ is a smooth function of $\vec{b}$ as compared to $\Gamma(\vec{b})$, at $A \gg 1$ we obtain [9]

$$\Phi(\vec{r}, \vec{r}') = \exp\left[-\frac{1}{2}\sigma_{tot}(pN)(1 - i\rho)t(\vec{b}, z) - \frac{1}{2}\sigma_{tot}(pN)(1 + i\rho)t(\vec{b}', z')\right.$$
$$\left. + \eta(\vec{b} - \vec{b}')\sigma_{el}(pN)t(\frac{1}{2}(\vec{b} + \vec{b}'), max(z, z'))\right]. \qquad (18)$$

Here the rapid dependence on $\vec{b} - \vec{b}'$ is concentrated in

$$\eta(\vec{b}) = \frac{\int d^2\vec{\Delta}\Gamma^*(\vec{b} - \vec{\Delta})\Gamma(\vec{\Delta})}{\int d^2\vec{\Delta}|\Gamma(\vec{\Delta})|^2} = \frac{1}{\pi\sigma_{el}(pN)}\int d^2\vec{q}\frac{d\sigma_{el}(pN)}{dq^2}\exp(i\vec{q}\vec{b}) = \exp\left[-\frac{\vec{b}^2}{4b_0^2}\right]. \qquad (19)$$

A comparison with the optical model DWIA is in order. Here one first constructs the phenomenological optical potential $V_{opt}(\vec{r})$ averaging over positions of spectator nucleons (for the review see [3]). Then, the wave equation for the struck proton wave is solved and produces the eikonal phase factor

$$S_{opt}(\vec{r}) = \exp\left[-\frac{i}{v}\int_z^\infty d\xi V_{opt}(\vec{b}, \xi)\right] \qquad (20)$$

( $v$ is the velocity of the struck proton) and factorized FSI factor

$$\Phi_{opt}(\vec{r},\vec{r}') = S_{opt}(\vec{r})S^*_{opt}(\vec{r}') . \tag{21}$$

When $V_{opt}(\vec{r})$ is constructed in the Glauber approximation, then

$$S^{(Gl)}_{opt}(\vec{r}) = \exp\left[-\frac{1}{2}\sigma_{tot}(pN)(1-i\alpha_{pN})t(\vec{b},z)\right] . \tag{22}$$

Compared to the Glauber theory, the optical model approximation (21) misses the $\propto \eta(\vec{b}-\vec{b}')$ term in the exponent of (18).

## 7  FSI in heavy nuclei: incoherent elastic rescatterings and transverse missing momentum distribution [9]

For heavy nuclei, one can generalize Eq. (13) and develop [9] a systematic multiple elastic rescattering expansion for $W_\perp(\vec{p}_\perp) = \int dp_{m,z} W(\vec{p}_m)$, which can be written in the form

$$W_\perp(\vec{p}_\perp) = \frac{1}{(2\pi)^2}\int d^2\vec{b}\,dz\,d^2\vec{\Delta}\exp(i\vec{p}_\perp\vec{\Delta})\rho(\vec{b}+\frac{1}{2}\vec{\Delta},z,\vec{b}-\frac{1}{2}\vec{\Delta},z)$$
$$\times S^{(Gl)}_{opt}(\vec{b}+\frac{1}{2}\vec{\Delta},z)S^{(Gl)\dagger}_{opt}(\vec{b}-\frac{1}{2}\vec{\Delta},z)\exp\left[\eta(\vec{\Delta})\sigma_{el}(pN)t(\vec{b},z)\right] . \tag{23}$$

To the zeroth order in $\eta(\vec{\Delta})$, Eq. (23) defines the local $\vec{p}_\perp$ distribution [9]

$$W_{\perp,opt}(\vec{b},z,\vec{p}_\perp) = \frac{1}{(2\pi)^2}\int d^2\vec{\Delta}\exp(i\vec{p}_\perp\vec{\Delta})\rho(\vec{b}+\frac{1}{2}\vec{\Delta},z,\vec{b}-\frac{1}{2}\vec{\Delta},z)$$
$$\times\rho_A^{-1}(\vec{b},z)S^{Gl\dagger}_{opt}(\vec{b}-\frac{1}{2}\vec{\Delta},z)S^{Gl}_{opt}(\vec{b}+\frac{1}{2}\vec{\Delta},z) , \tag{24}$$

which includes the FSI distortions at the optical level (22). This local distribution (24) is normalized as $\int d^2\vec{p}_\perp W_{\perp,opt}(\vec{b},z,\vec{p}_\perp) = |S^{Gl}_{opt}(\vec{b},z)|^2$. Expansion of the last exponential factor in Eq. (23) in powers of $\eta(\vec{\Delta})$ gives the multiple incoherent elastic rescattering series $W_\perp(\vec{p}_\perp) = \sum_{\nu=0}^{\infty} W_\perp^{(\nu)}(\vec{p}_\perp)$, where the zeroth order term is an averaged local $\vec{p}_\perp$ distribution,

$$W_\perp^{(0)}(\vec{p}_\perp) = \int d^2\vec{b}\,dz\,\rho_A(\vec{b},z)W_{\perp,opt}(\vec{b},z,\vec{p}_\perp) , \tag{25}$$

and the contribution of $\nu$-fold component for $\nu \geq 1$ reads

$$W_\perp^{(\nu)}(\vec{p}_\perp) = \frac{1}{\nu!}\int d^2\vec{b}\,dz\,\rho_A(\vec{b},z)t^\nu(\vec{b},z)\int\prod_{i=1}^{\nu}d^2\vec{q}_i\left(\frac{1}{\pi}\frac{d\sigma_{el}(pN)}{dq_i^2}\right)W_{\perp,opt}(\vec{b},z,\vec{p}_\perp-\sum_{j=1}^{\nu}\vec{q}_j) . \tag{26}$$

The convolution form of (26) suggest probabilistic reinterpretation, which is possible only in the $p_{m,z}$-integrated case, though. Distortions of the local distribution (24) are not small and Eqs. (24-26) of [9] do substantially improve the simplified version of multiple elastic rescattering expansion first derived in [4]. Putting $\vec{b} = \vec{b}'$ in the FSI factor (18) as it was done in several works [19], one misses the FSI contributions to the large-$p_\perp$ spectrum.

# 8 FSI in heavy nuclei: the longitudinal missing momentum distribution [9]

Next we consider the longitudinal missing momentum distribution

$$W_z(p_{m,z}) = \int d^2\vec{p}_\perp W(\vec{p}_m) = \frac{1}{2\pi} \int d^2\vec{b}\,dz\,dz'\, \rho(\vec{b},z,\vec{b},z')\Phi_z(\vec{b},z,z') \exp[i\vec{p}_{m,z}(z-z')]. \quad (27)$$

Here, the FSI factor equals $\Phi_z(\vec{b},z,z') = \Phi^{(in)}_{z,opt}(\vec{b},z,z')C_1(\vec{b},z,z')C_2(\vec{b},z,z')$, where [9]

$$\Phi^{(in)}_{z,opt}(\vec{b},z,z') = \exp\left[-\frac{1}{2}\sigma_{in}(pN)t(b,z) - \frac{1}{2}\sigma_{in}(pN)t(b,z'),\right] \quad (28)$$

$$C_1(\vec{b},z,z') = \exp\left[\frac{i}{2}\sigma_{tot}(pN)\alpha_{pN}\left(t(b,z) - t(b,z')\right)\right], \quad (29)$$

$$C_2(\vec{b},z,z') = \exp\left[-\frac{1}{2}\sigma_{el}(pN)\left|t(b,z) - t(b,z')\right|\right]. \quad (30)$$

The optical FSI factor $\Phi^{(in)}_{z,opt}(\vec{b},z,z')$ is a symmetric function of $z, z'$ and describes a distortion of the struck proton wave due to inelastic interactions (absorption) of the struck proton in a nucleus. In the approximation of $C_1 = C_2 = 1$, Eq. (27) defines the optical longitudinal missing momentum distribution $W^{(in)}_{z,opt}(p_{m,z})$, which is an even function of $p_{m,z}$. For the purposes of the qualitative analysis, for $A \gg 1$ useful approximations are $C_1(\vec{b},z,z') = \exp\left[-ik_1(z-z')\right]$ and $C_2(\vec{b},z,z') = \exp\left[-k_2|z-z'|\right]$, where

$$k_1 = \frac{1}{2}\sigma_{tot}(pN)\alpha_{pN}\langle n_A\rangle, \quad (31)$$

$$k_2 = \frac{1}{2}\sigma_{el}(pN)\langle n_A\rangle, \quad (32)$$

and $\langle n_A\rangle$ is the average nuclear density. Then, $W_z(p_{m,z})$ can be represented as [9]

$$W_z(p_{m,z}) \approx \frac{1}{2\pi}\int dk\, W^{(in)}_{z,opt}(p_{m,z} - k_1 - k)c_2(k), \quad (33)$$

where $c_2(k)$ stands for the Fourier transform of the factor $C_2$. The effect of the real part of the $pN$-amplitude leads, via the factor $C_1$, to an effective shift [5] of $p_{m,z}$, by $k_1 \sim 20$ MeV/c in the GeV's energy range. This shift produces a substantial forward-backward asymmetry.

In the CEBAF domain of $Q^2$, $k_2 \sim 10 - 20$ MeV/c $\ll k_F$ and $c_2(k)$ is a sharp function of $k$, much narrower than the conventional SPMD,

$$c_2(k) = \int d\xi \exp(ik\xi)\exp(-k_2|\xi|) = \frac{2k_2}{k^2 + k_2^2}. \quad (34)$$

At $|p_{m,z}| \lesssim k_F$ it acts like the $\delta$-function and $W_z(p_{m,z}) \approx W^{(in)}_{z,opt}(p_{m,z} - k_1)$. On the other hand, at a sufficiently large $|p_{m,z}| \gtrsim k_F$, the $p_{m,z}$ dependence of $W_z(p_{m,z})$ will be controlled rather by the slow asymptotic decrease of $c_2(k)$ with the resulting tail $W_z(p_{m,z}) \propto p_{m,z}^{-2}$. This asymptotics derives only from the $C_2$ being a non-analytical function of $z - z'$ and is not affected by the finite nucleus size, which changes somewhat the functional form of $c_2(k)$.

The integration over $\vec{r}, \vec{r}'$ in Eqs. (2,3) shows that the cross section of $A(e,e'p)$ scattering is a result of manifestly quantum interference of amplitudes with different locations of the struck proton. The factor $\theta(z_i - z')\theta(z_i - z)$ excludes the $\Gamma^\dagger \Gamma$ interaction over the part of either of the two trajectories in the calculation of the OBDM. This exclusion is of a purely quantum-mechanical origin and the $\propto p_{m,z}^{-2}$ tail of $W(\vec{p}_m)$ defies any classical interpretation. Evidently, this large-$p_{m,z}$ tail would be missed if the optical model FSI factor (22) were used (see also the above discussion of the large-$p_\perp$ tail). Following [9], we comment on what makes the Glauber theory and optical model DWIA descriptions of FSI different.

The Glauber theory distortion factor (4) also describes an eikonal solution of wave equation for the struck proton, but for a fixed configuration of spectator nucleons. Then, the reduced nuclear matrix element squared is computed, the sum over final states is performed and, finally, averaging over positions of spectator nucleon is performed. Evidently, a rigorous quantum-mechanical evaluation of the probability distribution for a subsystem (the struck proton in the considered case) for the process including a complex system (the struck proton and spectator nucleons in our case) requires calculations at the level of the density matrix of the subsystem. The Glauber theory equations Eqs. (2,3,16) embody precisely this procedure for the quasielastic $(e, e'p)$ scattering at high energy of the struck proton, when interaction with spectators can be treated in the fixed scatterer approximation. Because of the different order of operations as it was outlined in Section 6, the important contribution of $\Gamma^\dagger \Gamma$ interaction between the two trajectories in FSI-distorted OBDM is missed in the optical model DWIA.

For the quantitative discussion of FSI effects for heavy targets we refer to Ref. [9]. Here we only wish to comment on the so-called local density approximation (LDA)

$$\rho(\vec{r}, \vec{r}\,') = \rho_A(\frac{1}{2}(\vec{r} + \vec{r}\,'))W(\vec{r} - \vec{r}\,') \,, \qquad (35)$$

what is generally believed to be a good approximation for heavy nuclei. The results of Ref. [9] show this is not the case and the difference between $W(\vec{p}_m)$ obtained with the full shell model density matrix (15) and its LDA form (35) is substantial even for A=40 and even for small $p_m$.

## 9 An interplay of ISC and FSI effects: $^4He(e, e'p)$ as a testing ground [7,11]

The $^4He$ is a simple enough nucleus in which $W(\vec{p}_m)$ can be calculated accurately to all orders in the FSI and pair correlation function, although such a calculation is quite a formidable task [11]. For the high density of the $^4He$, an exhaustive analysis of FSI effects in the $^4He$ is a good guidance to a numerical significance of FSI effects in heavier nuclei.

The principal effects of ISC on SPMD were already reviewed in Section 4. Fig. 8 shows how the large-$p_m$ tail of SPMD builds up from the mean field approximation, $C_o = 0$, to soft core correlation $C_o = 0.5$ to hard core correlation $C_o = 1$. The destructive interference between the mean field (9) and the lowest order ISC (10) contributions is obvious at intermediate $p_m \sim k_F$. Remarkably, the extra strength at large $p_m$ due to short range correlations, comes from a depletion of SPMD at $p_m \sim k_F$, rather than from the region of $p_m \sim 0$, in which SPMD is rather enhanced by short range correlation effects [11].

Modulo to the different $p_\perp$ and $p_{m,z}$ dependence, the lowest order FSI contribution $W(\Gamma^\dagger + \Gamma; \vec{p}_m)$ is very similar to $N(C^\dagger + C; \vec{p}_m)$ of Eq. (9), but has larger normalization

$$\frac{W(\Gamma^\dagger + \Gamma; \vec{p}_m)}{N(C^\dagger + C; \vec{p}_m)} \sim \left(\frac{\sigma_{tot}}{4\pi b_0^2}\right) \cdot \left(\frac{b_0}{R_0}\right)^2 \cdot \left(\frac{R_0}{r_c}\right)^3 \cdot \frac{1}{C_0} \sim \frac{R_0}{r_c} \cdot \frac{1}{C_0} \gg 1 \,. \qquad (36)$$

The enhancement factor $\sim R_0/r_c$ in (36) derives from the fact that the Glauber operator is a long ranged function of the longitudinal separation in contrast to the short ranged correlation function. The driving short ranged FSI contribution to the OBDM and to the large-$p_m$ tail of $W(\vec{p}_m)$ comes from the $\propto \Gamma^\dagger(\vec{b}' - \vec{b}_i')\Gamma(\vec{b} - \vec{b}_i)$ terms in the expansion (7),

$$W(\Gamma^\dagger \Gamma; \vec{p}_m) \propto \left| \int d^2\vec{b}\, \Gamma(\vec{b}) \exp(i\vec{p}_\perp \vec{b}) \right|^2 = 4\pi \frac{d\sigma_{el}}{dp_\perp^2} = \frac{1}{4}\sigma_{tot}^2 (1+\rho^2) \exp(-b_o^2 p_\perp^2), \tag{37}$$

c.f. with (13) for the deuteron and (26) for heavy nuclei. (Wherever justified, we suppress the correction factors $[1 + \mathcal{O}(b_0/R_0)^2 \ll 1)]$.) Because of $b_0 \approx r_c$, the two leading large-$p_\perp$ contributions, $W(C^\dagger C; \vec{p}_m) = N(C^\dagger C; \vec{p}_m)$ and $W(\Gamma^\dagger \Gamma; \vec{p}_m)$, have a very similar $p_\perp$ dependence, but the normalization of the FSI term is much larger, c.f. Eq. (36),

$$\frac{W(\Gamma^\dagger \Gamma; \vec{p}_m)}{W(C^\dagger C; \vec{p}_m)} \approx \frac{1}{C_o^2 \sqrt{6}} \cdot \left[\frac{\sigma_{tot}}{4\pi r_c^2}\right]^2 \cdot \left(\frac{R_o}{r_c}\right)^2 \sim 7. \tag{38}$$

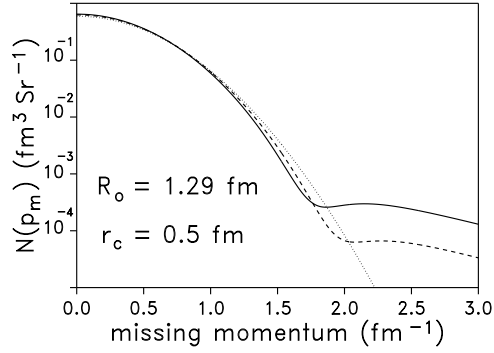

Figure 8: The PWIA momentum distribution $N(\vec{p}_m)$ in $^4He$ for the mean field approximation $C_o = 0$ (dotted curve), for soft core correlations $C_o = 0.5$ (dashed curve) and for hard core correlations $C_o = 1$ (solid curve) [11]. The other parameters are $R_o = 1.29 fm$ and $r_c = 0.5 fm$.

Consequently, the large-$p_\perp$ tail of $W(\vec{p}_m)$ must be completely dominated by FSI effects. A nontrivial quantum-mechanical effect of ISC-FSI interference due to the terms $\propto C^\dagger(\vec{r}' - \vec{r}_i')\Gamma(\vec{b} - \vec{b}_i), C(\vec{r} - \vec{r}_i)\Gamma^\dagger(\vec{b}' - \vec{b}_i')$ also contributes to a large-$p_\perp$ tail,

$$w(C\Gamma^\dagger + C^\dagger \Gamma; \vec{p}_m) \propto \int d^2\vec{r}\, C^\dagger(\vec{r}) \exp(i\vec{p}_\perp \vec{r}) \int d^2\vec{b}\, \Gamma(\vec{b}) \exp(-i\vec{p}_\perp \vec{b}) \propto \exp\left[-\frac{1}{2}(r_c^2 + b_o^2)p_\perp^2\right]. \tag{39}$$

The normalization of the FSI-ISC interference term contains the small factor $r_c/R_0$, but in the $^4He$ case this suppression is weak,

$$\frac{W(C\Gamma^\dagger + C^\dagger \Gamma; \vec{p}_m)}{W(\Gamma^\dagger \Gamma; \vec{p}_m)} \approx 4\sqrt{\frac{3}{5}} C_o \left(\frac{4\pi r_c^2}{\sigma_{tot}}\right) \cdot \frac{r_c}{R_o} \sim 1, \tag{40}$$

and the FSI-ISC interference effect is much more important than the pure ISC component $W(\vec{p}_m)$. Any semiclassical consideration would completely miss this large ISC-FSI interference effect. Now we present some of the results [11] for $W(\vec{p}_m)$. Unless specified otherwise, they are for hard core correlation, $C_0 = 1$.

Gross features of the departure of $W(\vec{p}_m)$ from the PWIA distribution $N(\vec{p}_m)$ for $^4He(e,e'p)$ are very similar to those in the $D(e,e'p)$ case. We proceed directly to the most interesting issue of whether one can probe the strength $C_0$ of short range correlations. The results shown in Fig. 9. demonstrate that, in a striking contrast to the strongly correlation dependent PWIA distribution $N(\vec{p}_m)$ of Fig. 8., the FSI-modified $W(\vec{p}_m)$ is *extremely insensitive* to the strength of short range correlation. Namely, in the transverse kinematics $W(\vec{p}_m)$ for hard core, $C_o = 1$, is only by $\sim 50\%$ larger than $W(\vec{p}_m)$ calculated with $C_o = 0$, in contrast to a difference of several orders of magnitude in the case of PWIA (see Fig.8). This enhancement is much stronger than the pure ISC contribution of PWIA and is quite counterintuitive, because naively one would expect that the hard core repulsion suppresses the classical probability of elastic rescattering of the struck proton on the spectator nucleon! The found enhancement of $W(\vec{p}_m)$ from the mean field, $C_0 = 0$, to hard core, $C_0 = 1$, case must be attributed to the above discussed ISC-FSI interference effect, see Eq. (39).

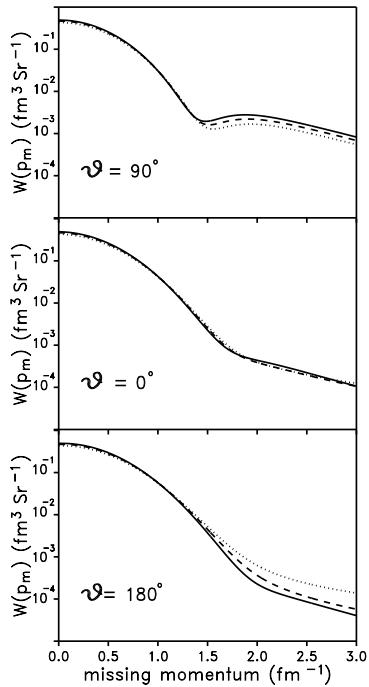

Figure 9: Predictions [11] for the FSI-modified missing momentum distribution $W(\vec{p}_m)$ in the $^4He(e,e'p)$ reaction are plotted for hard core correlations, $C_o = 1$, (solid line), soft core correlations, $C_o = 0.5$, (dashed line) and no correlations at all, $C_o = 0$ (dotted line), for different angles $\theta$. The upper panel shows the results for $\theta = 90°$, $\theta = 0°$ is shown in the middle panel and $\theta = 180°$ is shown in the lower panel. The other parameters are $R_o = 1.29 fm$ and $r_c = 0.5 fm$.

The situation in longitudinal kinematics is very tricky, as here one encounters a still another strong ISC-FSI correlation effect, this time connected with the real part of the $pN$ elastic scattering amplitude. In Fig. 10 we show $W_+(p_m) = \frac{1}{2}[W(\theta = 0°; p_m) + W(\theta = 180°)]$, which is free of the FSI contribution $\propto \rho$. In the mean field approximation, $C_0 = 0$, the large-$p_m$ tail of $W_+(p_m)$ is entirely due to the $\theta$-function effects and is numerically very close to the PWIA distribution at $C_0 = 1$. When both the FSI and ISC effects are simultaneously included, with the increase of $C_0$ the rising ISC contribution compensates partly for a suppression by short range repulsion of the $\theta$-function component of the FSI contribution. The net effect is a weak depletion of $W_+(p_m)$ from the mean field, $C_0 = 0$, value to the soft core correlation value at $C_0 = 0.5$. However, there is hardly any change in $W_+(p_m)$ from the soft core to hard core correlation result for $W_+(p_m)$.

The substantial rôle of the real part of the $pN$ scattering amplitude in this region of large

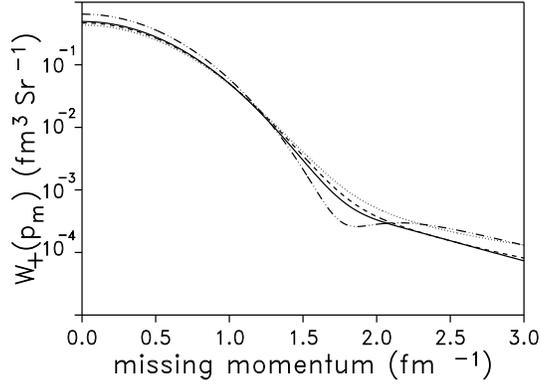

Figure 10: The predictions [11] for the FSI-modified distribution $W_+(p_m) = \frac{1}{2}[W(\theta = 0°; p_m) + W(\theta = 180°)]$ in the $^4He(e, e'p)$ reaction are shown for hard core correlations, $C_o = 1$, (solid line), soft core correlations, $C_o = 0.5$, (dashed line) and no correlations at all, $C_o = 0$, (dotted line). The other parameters are $R_o = 1.29 fm$ and $r_c = 0.5 fm$. For comparison, the PWIA momentum distribution $N(\vec{p}_m)$ for hard core correlations, $C_o = 1$, is also shown (dash-dotted line).

$p_m$ is obvious from the FSI-induced forward-backward asymmetry $A_{FB}(p_m)$, which is shown in Fig. 11. For soft and hard core correlations, $A_{FB}(p_m)$ for the $^4He$ target is remarkably similar to $A_{FB}(p_m)$ in $D(e, e'p)$ scattering shown in Fig. 3. In the deuteron case, the realistic wave functions [18] directly include the effects of short distance proton-neutron interaction. From this $^4He - D$ comparison we can conclude that, first, our simple Ansatz wave function correctly models gross features of short-distance nucleon-nucleon interaction in the $^4He$ and, second, the found change of the sign of $A_{FB}(p_m)$ and its rise with the correlation strength at large $p_m$ are on firm grounds. It is this enhancement of $A_{FB}(p_m)$ which effectively cancels the effect of slight decrease of $W_+(p_m)$ and produces the correlation independent $W(\theta = 0°; p_m)$. It is this enhancement of $A_{FB}(p_m)$ which amplifies the slight decrease of $W_+(p_m)$ and produces the counterintuitive substantial decrease of $W(\theta = 180°, p_m)$ with the correlation strength $C_0$.

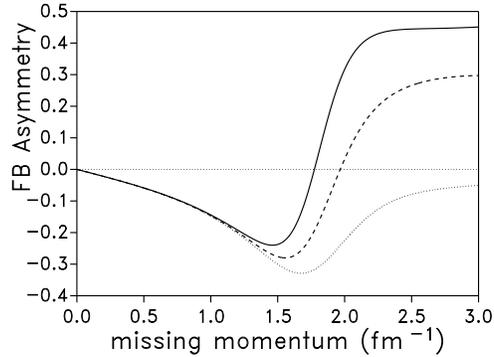

Figure 11: The predicted [11] forward-backward asymmetry $A_{FB}(p_m)$ is shown for $^4He$ for hard core correlations, $C_o = 1$, (solid line), soft core correlations, $C_o = 0.5$, (dashed line) and no correlations at all, $C_o = 0$, (dotted line). The other ground state parameters are $R_o = 1.29 fm$ and $r_c = 0.5 fm$.

# 10 Implications for nuclear transparency studies

Nuclear effects in $A(e,e'p)$ scattering are often discussed in terms of the transparency ratio $T_A(\vec{p}_m) = W(\vec{p}_m)/N(\vec{p}_m)$, shown for the $^4He$ target in Fig. 12 [11]. Evidently, FSI effects in $T_A$ do not reduce to an overall renormalization of the observed missing momentum distribution by nuclear attenuation factor. Only at $p_m \approx 0$ the found $\approx 24\%$ depletion can be interpreted as a pure nuclear attenuation effect; at larger $p_m$ the FSI distortions effects leads to $T_A(\vec{p}_m)$ which exhibits both much stronger depletion and "antishadowing" behaviour $T_A(\vec{p}_m) \gg 1$, see Fig. 12. In the (anti)parallel kinematics, nuclear transparency is very strongly affected by the real part of the $pN$ scattering amplitude, the effect of which can not be interpreted in terms of attenuation altogether. At moderate missing momenta, $p_m \lesssim 1.3\, fm^{-1}$, there is hardly any sensitivity to the correlation strength, see also Fig. 11. The sensitivity to short range correlations must be still weaker in heavier nuclei (for the related discussion see [20]).

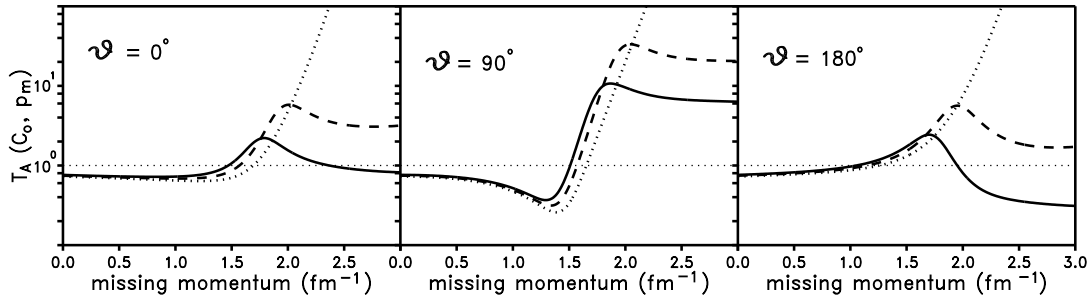

Figure 12: The nuclear transparency $T_A(\vec{p}_m) = \frac{W(C_0;\vec{p}_m)}{N(C_0;\vec{p}_m)}$ for $^4He(e,e'p)$ scattering is shown for hard core correlations, $C_o = 1$, (solid line), soft core correlations, $C_o = 0.5$, (dashed line) and no correlations at all, $C_o = 0$, (dotted line) [11]. The upper panel shows the results for $\theta = 90°$, $\theta = 0°$ is shown in the middle panel and $\theta = 180°$ is shown in the lower panel. The other ground state parameters are $R_o = 1.29 fm$ and $r_c = 0.5 fm$.

In the experimental determination of nuclear transparency one inevitably runs into a sort of vicious circle: The experimentally observed, FSI-distorted, $W(\vec{p}_m)$ must be compared to the SPMD $N(\vec{p}_m)$ which is not directly measurable. (Still further complications ensue if the measured cross section does not allow an integration over a sufficiently broad range of missing energy $E_m$.) The SPMD $N(\vec{p}_m)$ can only be calculated from certain models and can only partly be checked against the experimentally measured $W(\vec{p}_m)$, implicitly and/or explicitly assuming that the FSI effects can be factored out as an overall attenuation factor (for instance, see [13]). The above discussion clearly shows that one must be very careful, because in large parts of the phase space such an evaluation of nuclear transparency can lead astray.

# 11 FSI and multinucleon emission

In the quasi-deuteron picture, the large $p_m$ comes from interaction with the proton of the correlated pair. Simultaneously, the spectator nucleon of the correlated pair is ejected carrying the missing momentum $\vec{p}_m$, as it was discussed already in classic papers [1, 2]. Gross features of the contribution of FSI to multinucleon emission in $A(e,e'p)$ scattering were discussed in [4] with the conclusion that it is FSI which at large $p_m$ becomes a dominant source of multinucleon emission. To the discussion of large $p_\perp$ in [4] we must add that the strong FSI contribution to $W(\vec{p}_m)$ implies equally strong two-nucleon emission by FSI also in (anti)parallel kinematics. Therefore, backward ejection of spectator protons in $A(e,e'pp)$ scattering can not be taken for a clean signature of interaction with the correlated pair in a ground state of the target nucleus.

Evidently, elastic rescatterings of the struck proton lead to a broad peak in the missing energy spectrum at $E_m \sim p_m^2/2m_p$ [4]. The emergence of such a peak is quite obvious in transverse kinematics, but it persists in the longitudinal kinematics too. What shows that at large $p_m$ the missing energy spectrum extends to a very large $E_m$, and high $T_{kin}$ of the struck proton is crucial to exclude spurious effects from the $E_m$ dependence of FSI.

## 12 Discussion of the results and conclusions

We reviewed the recent progress in the theory of FSI in quasielastic $A(e,e'p)$ scattering [4]-[11]. We presented compelling evidence that for large-$p_m$ the observed missing momentum distribution is dominated by final state interaction of the struck proton with spectator nucleons and by the intricate interplay and quantal interference of FSI and ground state correlation effects. In transverse kinematics, the FSI contribution to $W(\vec{p}_m)$ exceeds the ISC contribution to the SPMD by the order of magnitude. Even here, a substantial part of the FSI effect comes from a quantum mechanical FSI-ISC interference effect in the one body density matrix, which defies a semiclassical interpretation. The pattern of FSI-ISC interference effects is still more complex for longitudinal kinematics, where we found a novel effect of strong enhancement of the forward-backward asymmetry by short range correlations in the ground state. In antiparallel kinematics, this ISC-FSI interference effect, which comes from the real part of the $pN$ scattering amplitude, leads to the FSI-modified $W(\vec{p}_m)$ which decreases with the correlation strength in the opposite to the SPMD. We are led to the conclusion that FSI effects **make impossible a model-independent determination of the SPMD $N(\vec{p}_m)$ and extraction of the short range correlation effects** from the experimentally measured missing momentum distribution $W(\vec{p}_m)$. Large FSI effects are of quite a general origin and are not an artifact of the Ansatz wave function used in our evaluations. We emphasize a simple and well understood origin of large enhancement parameters (36,38), which is a large radius of the nucleus as compared to a small radius of short range correlations.

One corollary of dominance of FSI effects at large $p_m$ scattering is a similarity of missing momentum spectra in $^2H(e,e'p)$ and $^4He(e,e'p)$ reactions (scaled up by the factor $\sim 3$ for the deuteron). Such a similarity emerges not because of the quasi-deuteron mechanism in the $^4He$, but because of the universality of final state proton-nucleon interaction in both nuclei [21].

**Acknowledgments:** N.N.N. thanks C.Ciofi degli Atti, S.Boffi and M.Giannini for the invitation to the exciting Conference on Perspectives in Nuclear Physics at Intermediate Energies, held in the stimulating environment of the International Centre for Theoretical Physics.

## References

[1] K.Gottfried, *Ann.Phys.* **21** (1963) 29; W.Czyż and K.Gottfried, *Nucl.Phys.* **21** (1961) 676; *Ann.Phys.* **21** (1963) 47.

[2] N.Srivastava, *Phys.Rev.* **135B** (1964) 612; D.U.L.Yu, *Ann. Phys. (NY)* **38** (1966) 392; J.W.Van Orden, W.Truex and M.K.Banerjee, *Phys. Rev.* **C21** (1980) 2628.

[3] S.Frullani and J.Mougey, *Adv. Nucl. Phys.*, Editors J.W.Negele and E.Vogt **14**, 3 (1984); A.E.I.Dieperink and P.K.A. de Witt Huberts, *Annu. Rev. Nucl. Part. Sci.* **40** (1990) 239; S.Boffi, C.Giusti and F.D.Pacati, *Phys.Rep.* **226** (1993) 1.

[4] N.N.Nikolaev, A.Szczurek, J.Speth, J.Wambach, B.G.Zakharov and V.R.Zoller, *Nucl.Phys* **A582** (1995) 665.

[5] N.N.Nikolaev, A.Szczurek, J.Speth, J.Wambach, B.G.Zakharov and V.R.Zoller, *Phys.Rev.* **C50** (1994) R1296.

[6] J.Nemchik, N.N.Nikolaev and B.G.Zakharov, Proceedings of the Workshop on CEBAF at Higher Energies, CEBAF, April 14-16, 1994, Editors: N.Isgur and P.Stoler, pp. 415-464.

[7] A.Bianconi, S.Jeschonnek, N.N.Nikolaev and B.G.Zakharov, *Phys. Lett.* **B338** (1994) 123.

[8] A.Bianconi, S.Jeschonnek, N.N.Nikolaev and B.G.Zakharov, *Phys. Lett.* **B343** (1995) 13.

[9] N.N.Nikolaev, J.Speth and B.G.Zakharov, Jülich preprint **KFA-IKP(Th)-1995-01** (1995), submitted to *Nucl. Phys.* **A**.

[10] A.Bianconi, S.Jeschonnek, N.N.Nikolaev and B.G.Zakharov, Jülich preprint **KFA-IKP(Th)-1995-02** (1995), submitted to *Phys. Rev. C*.

[11] A.Bianconi, S.Jeschonnek, N.N.Nikolaev and B.G.Zakharov, Jülich preprint **KFA-IKP(Th)-1995-13**.

[12] R.J.Glauber, in: *Lectures in Theoretical Physics*, v.1, ed. W.Brittain and L.G.Dunham. Interscience Publ., N.Y., 1959; R.J.Glauber and G.Matthiae, *Nucl. Phys.* **B21** (1970) 135.

[13] N.C.R.Makins et al., *Phys. Rev. Lett* **72** (1994) 1986; T.G. O'Neill et al., *Phys. Lett.* **B351** (1995) 93; N.C.R.Makins and R.G.Milner, *preprint MIT-LNS* **94-79**.

[14] T. de Forest Jr., *Nucl.Phys.* **A392** (1983) 232.

[15] R.I.Dzhibuti and R.Ya.Kezerashvili, *Sov. J. Nucl. Phys.* **20** (1974) 17; M.Traini and G.Orlandini, *Z. Phys.* **A321** (1985) 479; G.Có, A.Fabrocini and S.Fantoni, *Nucl. Phys.* **A568** (1994) 73.

[16] G.D.Alkhazov, S.I.Belostotsky and A.A.Vorobyev, *Phys. Rep.* **C42** (1978) 89.

[17] T.Lasinski et al., *Nucl. Phys.* **B37** (1972) 1; C.Lechanoine-LeLuc and F.Lehar, *Rev. Mod. Phys.* **65** (1993) 47.

[18] R.Machleidt, K.Holinde, and C.Elster, *Phys. Rep.* **149** (1987) 1; M.Lacombe, B.Loiseau, R.Vinh Mau, J.Cote, P.Pires and R.de Tourreil, *Phys. Lett.* **B101** (1981) 139.

[19] A.Kohama, K.Yazaki and R.Seki, *Nucl. Phys.* **A551** (1993) 687; A.S.Rinat and B.K.Jennings, *Nucl. Phys.* **A568** (1994) 873.

[20] N.N.Nikolaev, A.Szczurek, J.Speth, J.Wambach, B.G.Zakharov and V.R.Zoller, *Phys. Lett.* **B317** (1993) 281.

[21] A.Bianconi, S.Jeschonnek, N.N.Nikolaev and B.G.Zakharov, Jülich preprint **KFA-IKP(Th)-1995-14** (1995).